\documentclass[aps,prx,twocolumn,notitlepage,10pt,longbibliography]{revtex4-2}
\usepackage{graphicx}
\usepackage{amsfonts,amsmath,amssymb}
\usepackage[colorlinks=true]{hyperref}
\newcommand{\aaa}{\mathbf a}

\newcommand{\mmm}{\mathbf m}

\begin{document}
\title{Helical edge modes in a triangular Heisenberg antiferromagnet}
\author{Basti{\'a}n Pradenas}
\author{Grigor Adamyan}
\author{Oleg Tchernyshyov}
\affiliation{
William H. Miller III Department of Physics and Astronomy, 
Johns Hopkins University,
Baltimore, Maryland 21218, USA
}
\date{\today}

\begin{abstract}

We investigate the emergence of helical edge modes in a Heisenberg antiferromagnet on a triangular lattice, driven by a topological mechanism similar to that proposed by Dong \emph{et al.} [Phys. Rev. Lett. \textbf{130}, 206701 (2023)] for chiral spin waves in ferromagnets. The spin-frame field theory of a three-sublattice antiferromagnet allows for a topological term in the energy that modifies the boundary conditions for certain polarizations of spin waves and gives rise to edge modes. These edge modes are helical: modes with left and right circular polarizations propagate in opposite directions along the boundary in a way reminiscent of the electron edge modes in two-dimensional topological insulators. The field-theoretic arguments are verified in a realistic lattice model of a Heisenberg antiferromagnet with superexchange interactions that exhibits helical edge modes. The strength of the topological term is proportional to the disparity between two inequivalent superexchange paths. These findings suggest potential avenues for realizing magnonic edge states in frustrated antiferromagnets without requiring Dzyaloshinskii-Moriya interactions or nontrivial magnon band topology.
  
\end{abstract}

\maketitle

\section{Introduction}
\label{sec:intro}

Chiral edge states first attracted the attention of physicists in connection with the integer quantum Hall effect. Electrons travel in only one direction along the edge of a quantum Hall sample with an integer filling of its Landau levels \cite{Halperin:1982}. The edge separates two distinct types of insulator---the quantum Hall state and the vacuum---which have different values of a topological invariant (the net Berry flux of the filled electron bands). Because a topological invariant cannot be changed continuously, a spatial transition between the two topologically distinct insulators must go through a conducting state, hence the metallic edge mode. The one-way character of mode propagation along the edge requires the breaking of the time-reversal symmetry, which, in the case of the quantum Hall effect, is provided by a strong applied magnetic field. 

Similar edge states are also possible when time reversal is not broken. In two-dimensional topological insulators, electrons with spin up travel one way along the edge and electrons with spin down the other way. Such states are called \emph{helical}, in reference to the helicity of a neutrino, whose spin is locked to its travel direction \cite{Hasan:2010}. 

Other quantum particles, not just electrons, can have chiral or helical edge states. Photons---quanta of light---form chiral edge modes in metamaterials with topologically nontrivial photonic band structures \cite{Haldane:2008, Wang:2009}. Helical edge modes for magnons---quanta of spin waves---have also been predicted theoretically \cite{Matsumoto:2011, Shindou:2013, Zhang:2013}. In this case, a topologically nontrivial band structure arises in the presence of the Dzyaloshinskii-Moriya interaction. 

Chiral and helical edge states typically live in the gaps separating bulk energy bands. The fermionic statistics of electrons makes it possible to fill an integer number of bulk bands. With the Fermi level between two bulk bands, a chiral edge band connecting these bulk bands must cross the Fermi level. Thus, low-energy electronic excitations are confined to these chiral edge modes. In contrast, photons and magnons are bosonic particles, for which there is no analog of band filling. For this reason, photonic and magnonic chiral edge states are usually high-energy excitations. 

An entirely different mechanism for the formation of magnonic chiral edge modes was  recently proposed by \textcite{Dong:2023}. They considered a two-dimensional model of a ferromagnet where magnetization arises in part from orbital motion of conduction electrons and in part from spin. When a conduction electron is moving through a spin texture described by a unit-vector field $\mmm(x,y)$ parallel to the local direction of spin, the locking of the conduction electron's spin to the direction of $\mmm(x,y)$ creates a geometric (Berry) phase whose curvature produces an effect similar to that of a magnetic field of strength $B_\text{geom} = \frac{\Phi_0}{4\pi} \, \mmm \cdot (\partial_x \mmm \times \partial_y \mmm)$, where $\Phi_0 = hc/e$ is the magnetic flux quantum. The interaction of the orbital magnetization $M_\text{orb}$ and its spin counterpart has the form of the Zeeman coupling,
\begin{equation}
U_\text{geom} = - \int dx \, dy \, M_\text{orb} B_\text{geom} 
= - M_\text{orb} \Phi_0 Q,
\label{eq:U-geom}
\end{equation}
where 
\begin{equation}
Q = \frac{1}{4\pi} 
\int dx \, dy \, 
\mmm \cdot (\partial_x \mmm \times \partial_y \mmm)    
\label{eq:m-skyrmion-number}
\end{equation}
is known as the skyrmion number. 

The energy term \eqref{eq:U-geom} is topological in nature. If the spin texture is localized and $\mmm(x,y)$ becomes uniform at spatial infinity, then $Q$ is an integer. Continuous variations of $\mmm(x,y)$ cannot change an integer. Therefore, the functional derivative of the energy \eqref{eq:U-geom} vanishes and has no effect on the classical dynamics of the spin field $\mmm(x,y)$. However, in a finite sample, $Q$ is not quantized, so the functional derivative $\delta U_\text{geom}/\delta \mmm$ need not vanish, and indeed it has a nonzero value at the boundary. This modifies the boundary conditions for the spin field and thus creates chiral spin-wave modes traveling along the boundary \cite{Dong:2023}. In contrast to magnonic edge modes induced by a nontrivial topology of the magnon wavefunctions in momentum space, this edge mode lies below the bulk band and is thus a low-energy spin excitation. 

In this paper, we describe a helical analog of the chiral edge mode of \textcite{Dong:2023} in a Heisenberg antiferromagnet on a triangular lattice. We have recently presented a general field theory of the Heisenberg antiferromagnet with three magnetic sublattices \cite{Pradenas:2024}. The first hint of a possible existence of helical edge modes in that field theory was found in the form of a topological term in the energy, similar to Eq.~\eqref{eq:U-geom}. Such a term is allowed for a Heisenberg antiferromagnet on a triangular lattice and indeed gives rise to helical spin-wave edge modes. We also found a realistic lattice model whose continuum limit harbors the topological term. Finally, we checked that the spin-wave spectrum of the lattice model contains helical spin-wave edge modes.  The edge modes reside below the bulk modes of the  antiferromagnet and thus represent low-energy excitations. 

The layout of the paper is as follows. In Sec.~\ref{sec:field-theory}, we present a field-theoretic argument for the possibility of helical magnon edge modes in a triangular Heisenberg antiferromagnet and derive the properties of the edge modes in the long-wavelength limit. Sec.~\ref{sec:lattice-model} describes a Heisenberg lattice model that yields the topological term enabling the edge modes. The field-theoretic treatment is refined to include trivial edge modes present in the absence of the topological term. Sec.~\ref{sec:nonlinear} contains a fully nonlinear analysis of the spin-wave edge modes. We conclude with a discussion in Sec.~\ref{sec:discussion}. 

\section{Field theory}
\label{sec:field-theory}

\subsection{Field theory of a triangular antiferromagnet}

\begin{figure}    \includegraphics[width=0.75\columnwidth]{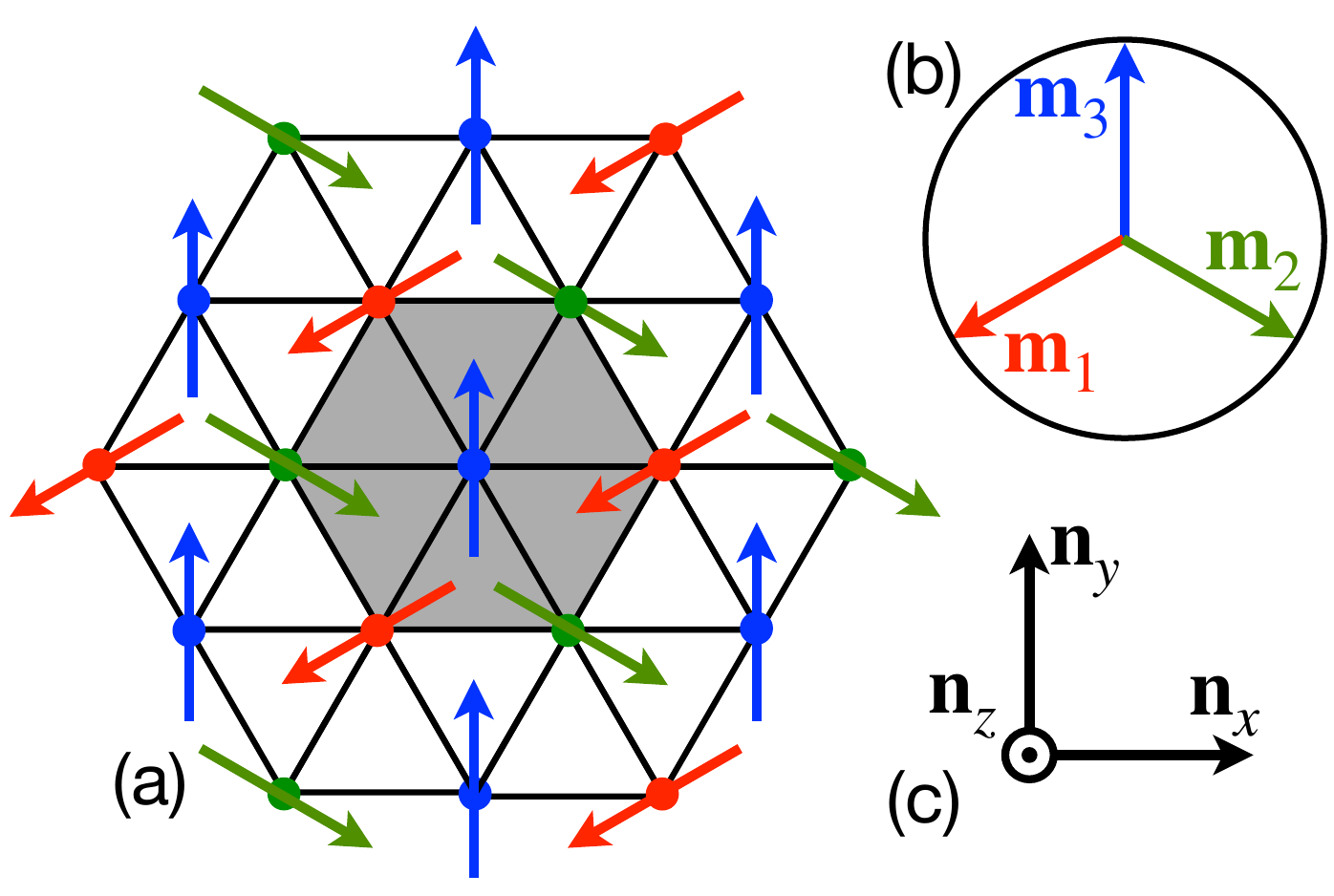}
\caption{(a) Magnetic order in a Heisenberg antiferromagnet on a triangular lattice. The shaded area is a magnetic unit cell. (b) The magnetic order parameter. (c) The corresponding spin-frame vectors.}
\label{fig:spin-frame}
\end{figure}

A Heisenberg antiferromagnet with a triangular lattice motif exhibits strong geometric frustration, forming a non-collinear magnetic order with three magnetic sublattices. In a ground state, the three spins of a triangle satisfy the relation $\mathbf S_1 + \mathbf S_2 + \mathbf S_3 = 0$, which implies that they are coplanar and make angles of $120^\circ$ with one another, Fig.~\ref{fig:spin-frame}(a). The magnetic order parameter can be thought of as a rigid body formed by unit vectors $\mathbf{m}_{1}$, $\mathbf{m}_{2}$, and $\mathbf{m}_{3}$, Fig.~\ref{fig:spin-frame}(b), that can be rotated as a whole without changing its shape. Because the orientation of a rigid body can be parametrized by an $SO(3)$ matrix, \textcite{Dombre:1989} formulated a matrix $\sigma$ model of a triangular-lattice antiferromagnet. An alternative description, using an orthogonal spin frame as the order parameter, was recently proposed by two of us \cite{Pradenas:2024}. To define the spin frame, one makes linear combinations of the three sublattice magnetization fields to form a uniform component of magnetization $\mathbf m$ and two staggered components $\mathbf n_x$ and $\mathbf n_y$:
\begin{equation}
\begin{split}
\mathbf{m}  & =
\mathbf{m}_1+\mathbf{m}_2+\mathbf{m}_3,
\\
\mathbf{n}_x & = 
(\mathbf{m}_2-\mathbf{m}_1)/\sqrt{3}, \\
\mathbf{n}_y & = 
(2 \mathbf{m}_3-\mathbf{m}_2-\mathbf{m}_1)/3.
\end{split}
\label{eq:m-nx-ny-def}
\end{equation}
To these, we add the vector spin chirality 
\begin{equation}
\mathbf{n}_z=\frac{2(\mathbf{m}_1 \times \mathbf{m}_2+\mathbf{m}_2 \times \mathbf{m}_3+\mathbf{m}_3 \times \mathbf{m}_1)}{3 \sqrt{3}}.
\end{equation}
In a ground state, $\mathbf{m} = 0$; the three spin-frame vectors form an orthonormal frame, Fig.~\ref{fig:spin-frame}(b): 
\begin{equation}
    \mathbf{n}_i \cdot \mathbf{n}_j = \delta_{i j}, \quad \mathbf{n}_i \times \mathbf{n}_j = \epsilon_{i j k} \mathbf{n}_k.
\label{eq:spin-frame-orthonormality}
\end{equation}
The Latin subscripts take on the values $i = x$, $y$, and $z$; summation is implied over doubly repeated indices. The spin-frame vectors form a convenient basis for expressing spin-related vectors. 

The Cartesian subscripts of $\mathbf n_i$ indicate that the spin-frame vectors transform under the point-group symmetries of the lattice in the same way as do Cartesian components $C_i$ of a spatial vector. These symmetries restrict the energy density of exchange interactions to a superposition of three invariants \cite{Pradenas:2024},
\begin{equation}\label{eq:conjecture-exchange}
\mathcal{U}  
   = \frac{\lambda}{2}   \partial_{\alpha} \mathbf{n}_{\alpha} \cdot  \partial_{\beta} \mathbf{n}_{\beta}
   +  \frac{\mu}{2}\partial_{\alpha} \mathbf{n}_{\beta} \cdot \partial_{\alpha} \mathbf{n}_{\beta}     
   +\frac{\nu}{2}\partial_{\alpha} \mathbf{n}_{\beta} \cdot  \partial_{\beta} \mathbf{n}_{\alpha}.
\end{equation}
The Greek subscripts take on the values $x$ and $y$; summation is implied over doubly repeated indices. There are three spin-wave branches with linear spectra $\omega_n(k) = c_n k$ with the characteristic velocities $c$ defined by
\begin{equation}
c_\text{I}^2 ={\frac{\mu}{\rho}}, 
\quad
c_\text{II}^2 = {\frac{\lambda + \mu + \nu}{\rho}}, 
\quad
c_\text{III}^2 = {\frac{\lambda + 2\mu + \nu}{\rho}},
\label{eq:spin-wave-velocities-general}
\end{equation}
where $\rho$ is the inertia density \cite{Pradenas:2024}. 

It is worth noting that the first and third terms in the exchange energy density \eqref{eq:conjecture-exchange} are related through integration by parts. As a result of that, the stiffness coefficients $\lambda$ and $\nu$ enter the bulk equations of motion solely as the sum $\lambda+\nu$ \cite{Pradenas:2024}. In an antiferromagnet on a triangular lattice, translational symmetry of the lattice demands that these terms vanish in the bulk, so that $\lambda+\nu = 0$ and the bulk energy density of a Heisenberg antiferromagnet on a triangular lattice boils down to 
\begin{equation}
\mathcal U_\text{bulk} = 
\frac{\mu}{2}
\partial_{\alpha} \mathbf{n}_{\beta} 
\cdot 
\partial_{\alpha} \mathbf{n}_{\beta}.
\end{equation}
The spin-wave velocities \eqref{eq:spin-wave-velocities-general} become degenerate for two out of three branches: 
\begin{equation}
c_\text{I}^2 = c_\text{II}^2 = \frac{\mu}{\rho}, 
\quad
c_\text{III}^2 = \frac{2\mu}{\rho}.
\label{eq:spin-wave-velocities-triangular}
\end{equation}

This still leaves the possibility of having the first and third terms in Eq.~\eqref{eq:conjecture-exchange} with $\lambda = -\nu$. This combination has no effect in the bulk but produces interesting edge effects, as will be seen shortly. We thus focus our attention on the energy density term
proportional to 
\begin{equation}
\mathcal F_{xy}
= \frac{1}{2} 
(\partial_{\alpha} \mathbf{n}_{\alpha} 
\cdot  \partial_{\beta} \mathbf{n}_{\beta}
- \partial_{\beta} \mathbf{n}_{\alpha} 
\cdot \partial_{\alpha} \mathbf{n}_{\beta}).
\label{eq:F-xy-z-aabb-baab}
\end{equation}
For lack of a better name, we shall refer to this quantity as simply a \emph{topological density}. 

\subsection{Topological density}

To appreciate the potential significance of the topological density \eqref{eq:F-xy-z-aabb-baab}, we explore its alternative representations. 

It is straightforward to write down $\mathcal F_{xy}$ explicitly in terms of the $x$ and $y$ labels: 
\begin{equation}
\mathcal F_{xy} = 
\partial_{x} \mathbf{n}_{x} \cdot \partial_{y} \mathbf{n}_{y} 
-\partial_{y} \mathbf{n}_{x} \cdot \partial_{x} \mathbf{n}_{y}.   
\label{eq:F-xy-z-xxyy-yxxy}
\end{equation}
It can now be recast as the curl of a vector potential,
\begin{equation}
\begin{split}
\mathcal F_{xy} 
&= \partial_x \mathcal A_y 
- \partial_y \mathcal A_x, 
\\
\mathcal A_\alpha 
&\equiv \mathbf n_x \cdot 
\partial_\alpha \mathbf n_y
= - \mathbf n_y \cdot 
\partial_\alpha \mathbf n_x.
\end{split}
\label{eq-F-xy-z-curl-A}
\end{equation}
The last identity follows from the orthogonality of the spin-frame vectors \eqref{eq:spin-frame-orthonormality}. 

By Stokes' theorem, an area integral over the topological density $\mathcal F_{xy}$ reduces to a boundary integral of the vector potential:
\begin{equation}
\int_\Sigma d^2 r \, \mathcal F_{xy} 
= \oint_{\partial \Sigma} dx^\alpha \, \mathcal A_\alpha.
\end{equation}
Here $\partial \Sigma$ is the boundary of area $\Sigma$. The reduction of the area integral to a boundary one confirms our intuition expressed earlier: an energy term based on the area integral of the topological density \eqref{eq:F-xy-z-aabb-baab} is effective at the edge of the system, rather than in the bulk.

Another useful observation is the invariance of $\mathcal F_{xy}$ under (position-independent) rotations of the spin frame about the $\mathbf n_z$ axis. For an infinitesmimal rotation angle $\delta \theta_z$, we have $\delta \mathbf n_x = \mathbf n_y \delta \theta_z$ and $\delta \mathbf n_y = - \mathbf n_x \delta \theta_z$. This invariance suggests to us that perhaps $\mathcal F_{xy}$ can be expressed in terms of spin frame vector $\mathbf n_z$ and its derivatives. That is indeed the case: 
\begin{equation}
\mathcal F_{xy} 
= \mathbf n_z \cdot 
(\partial_x \mathbf n_z \times \partial_y \mathbf n_z). 
\label{eq:F-xy-z-skyrmion}
\end{equation}
The equivalence of expressions \eqref{eq:F-xy-z-xxyy-yxxy} and \eqref{eq:F-xy-z-skyrmion} can be verified directly by substituting $\mathbf n_z = \mathbf n_x \times \mathbf n_y$ into the latter. 

Integration of the topological density \eqref{eq:F-xy-z-skyrmion} over an infinite area yields a quantized result: 
\begin{equation}
\int_{\mathbb R^2} d^2 r \, 
\mathbf n_z \cdot 
(\partial_x \mathbf n_z \times \partial_y \mathbf n_z)
= 4 \pi Q,
\end{equation}
where $Q \in \mathbb Z$ is an integer-valued winding number counting how many times the plane $\mathbf R^2$ maps onto the unit sphere $S^2$. It is a topological invariant similar to the skyrmion number in a ferromagnet \eqref{eq:m-skyrmion-number}. Like \textcite{Dong:2023} for their model of an orbital ferromagnet, we expect to find spin waves confined to the sample edge. 

The mathematics of the spin frame has its analog in the theory of the superfluid phase A of $^3$He. The counterpart of vector chirality $\mathbf n_z$ is the orbital angular momentum $\mathbf l$ of a Cooper pair. The equivalence of expressions \eqref{eq-F-xy-z-curl-A} and \eqref{eq:F-xy-z-skyrmion} is the analog of the Mermin--Ho relation \cite{Mermin:1976}. 

\subsection{Minimal continuum model}

Our basic continuum model of a triangular Heisenberg antiferromagnet with a topological term has the energy density \eqref{eq:conjecture-exchange} of a generic 3-sublattice antiferromagnet with 
\begin{equation}
\lambda = - \nu = \mu \Delta.
\label{eq:lambda-nu-via-mu-Delta}
\end{equation}
The dimensionless parameter $\Delta$ expresses the strength of the topological edge term relative to the bulk term. The bulk and edge energies are
\begin{equation}
\begin{split}
U_\text{bulk} &= \frac{\mu}{2}
\int_\Omega d^2 r \,
\partial_{\alpha} \mathbf{n}_{\beta} \cdot \partial_{\alpha} \mathbf{n}_{\beta},
\\
U_\text{edge} &= \mu \Delta 
\int_\Omega d^2 r \, \mathcal F_{xy} 
= \mu \Delta \oint_{\partial \Omega}
dx^\alpha\, 
\mathbf n_x \cdot \partial_\alpha \mathbf n_y.
\end{split}
\label{eq:minimal-model}
\end{equation}
Here $\Omega$ is the sample area and $\partial \Omega$ is its boundary. 

To describe the dynamics of the antiferromagnet, we must add the kinetic energy of the spin frame \cite{Pradenas:2024},
\begin{equation}
K = \frac{\rho}{4}
\int_\Omega d^2 r \,
\partial_t \mathbf{n}_i \cdot \partial_t \mathbf{n}_i.
\end{equation}
Latin indices $i$ run through all three Cartesian labels, $i = x, y, z$, whereas Greek ones only through the first two, $\alpha = x, y$. 

\subsection{Linear spin waves}

To parametrize small-amplitude spin waves, it is convenient to introduce a vector field $\boldsymbol \theta$ encoding the local angle of rotation:
\begin{equation}
\mathbf n_i 
= \mathbf e_i 
+ \boldsymbol \theta \times \mathbf e_i
+ \frac{1}{2} \boldsymbol \theta \times 
(\boldsymbol \theta \times \mathbf e_i)
+ \ldots
\end{equation}
Here $\mathbf e_i$ are a triple of orthonormal vectors representing a reference ground state. Expressing the kinetic and potential energies in terms of the angle field and keeping terms up to the second order in $\theta_i = \boldsymbol \theta \cdot \mathbf e_i$ yields 
\begin{equation}
\begin{split}
K &= 
\int_\Omega d^2 r \,
\frac{\rho}{2}
\left[
(\partial_t \theta_x)^2 
+ (\partial_t \theta_y)^2 
+ (\partial_t \theta_z)^2
\right],
\\
U_\text{bulk} &= 
\int_\Omega d^2r \,
\frac{\mu}{2}
\left[
(\nabla \theta_x)^2 
+ (\nabla \theta_y)^2 
+ 2(\nabla \theta_z)^2
\right],
\\
U_\text{edge} &= 
\oint_{\partial \Omega} dx^\alpha \, 
\frac{\mu \Delta}{2} \,
(
\theta_x \partial_\alpha \theta_y 
- \theta_y \partial_\alpha \theta_x 
- 2\partial_\alpha \theta_z
).
\end{split}
\label{eq:K-U-phi}
\end{equation} 
Minimizing the action $S = \int dt \, (K - U)$ yields equations of motion in the bulk, 
\begin{equation}
\begin{split}
\rho \partial_t^2 \theta_x - \mu \nabla^2 \theta_x = 0, 
\\
\rho \partial_t^2 \theta_y - \mu \nabla^2 \theta_y = 0,
\\
\rho \partial_t^2 \theta_z - 2\mu \nabla^2 \theta_z = 0,
\end{split}
\label{eq:spin-waves-eom}
\end{equation}
and boundary conditions at the edge. With the $x$-axis parallel to the boundary, 
\begin{equation}
\begin{split}
\partial_y \theta_x - \Delta \partial_x \theta_y = 0,
\\
\partial_y \theta_y + \Delta \partial_x \theta_x = 0,
\\
\partial_y \theta_z = 0.
\end{split}
\label{eq:spin-waves-boundary}
\end{equation}
The bulk equations of motion \eqref{eq:spin-waves-eom} yield the anticipated spin-wave velocities \eqref{eq:spin-wave-velocities-triangular}. 

The topological term mixes the degenerate modes $\theta_x$ and $\theta_y$ through the boundary conditions \eqref{eq:spin-waves-boundary}. It is convenient to introduce two complex fields
\begin{equation}
\theta_\pm = \theta_x \pm i \theta_y.   
\label{eq:theta-pm}
\end{equation}
They have the same bulk equations of motion as $\theta_x$ and $\theta_y$, 
\begin{equation}
\rho \partial_t^2 \theta_\pm - \mu \nabla^2 \theta_\pm = 0, 
\label{eq:spin-waves-eom-pm}
\end{equation}
and have decoupled boundary conditions,
\begin{equation}
\partial_y \theta_\pm 
\pm i \Delta \partial_x \theta_\pm
= 0.
\label{eq:spin-waves-boundary-pm}
\end{equation}

\subsection{Helical edge modes}
\label{sec:field-theory:helical-modes}

Assuming that the magnet is in the infinite semi-plane $y>0$ with the boundary at $y=0$, we try the Ansatz
\begin{equation}
\theta_\pm(x,y,t) = \Theta e^{- i \omega t + i k x - \kappa y}.
\label{eq:edge-mode-ansatz}
\end{equation}
The spin frame is tilted by angle $\Theta e^{- \kappa y}$ about the rotating axis 
\begin{equation}
\mathbf n_\pm(t) = 
\mathbf n_x \cos{\omega t} \mp \mathbf n_y \sin{\omega t}.
\end{equation}
Thus modes $\theta_+$ and $\theta_-$ can be thought of as spin waves with the clockwise and counterclockwise circular polarizations. These waves propagate along the edge ($x$-axis) and decay into the bulk ($y$-axis). The inverse localization length $\kappa>0$ is related to the propagation wavenumber $k$ through the boundary conditions \eqref{eq:spin-waves-boundary-pm}: 
\begin{equation}
\kappa \pm \Delta k = 0.
\label{eq:kappa-Delta-k-linear}
\end{equation}
An exponentially decaying solution, $\kappa>0$, exists only for one sign of the wavenumber, 
\begin{equation}
\pm \Delta k < 0.    
\end{equation}
For the opposite sign of $k$, this mode increases exponentially with $y$ and is thus localized at the opposite edge. The frequency of the edge mode is determined by the bulk equation of motion \eqref{eq:spin-waves-eom-pm}: 
\begin{equation}
\omega^2 
= c_\text{I}^2(k^2 - \kappa^2) = c_e^2 k^2.    
\end{equation}
Edge waves propagate at a lower speed $c_e$ than its bulk counterparts:
\begin{equation}
c_e^2 = c_\text{I}^2(1 - \Delta^2).  
\label{eq:edge-mode-speed-infinitesimal}
\end{equation}

\begin{figure}
\includegraphics[width=0.95\columnwidth]{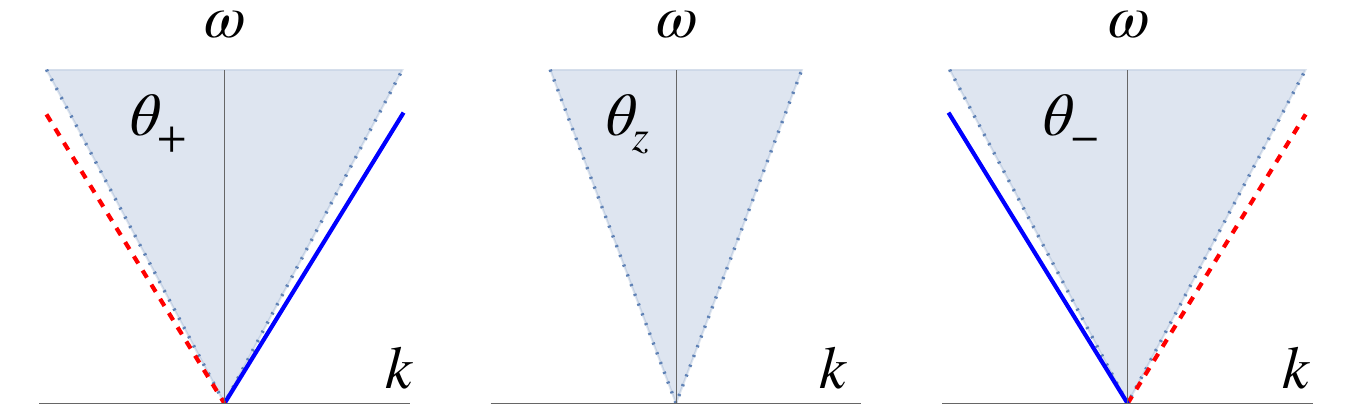}
\caption{Spin-wave spectra $\omega(k)$ for modes $\theta_\pm$ and $\theta_z$ in a strip for the minimal continuum model at $\Delta = 0.5$. Shaded area represents bulk modes; lines represent edge modes for the upper boundary (blue solid line) and lower boundary (red dashed line).}
\label{fig:strip-spectrum-minimal-model}
\end{figure}

The spin-wave spectrum $\omega(k)$ for our minimal model with $\Delta = 0.5$ in a wide strip is shown in Fig.~\ref{fig:strip-spectrum-minimal-model}. The $\theta_z$ mode has only bulk modes, whereas both $\theta_\pm$ modes have one mode propagating either right or left on each edge. Considered alone, edge waves of the $\theta_+$ and $\theta_-$ modes can be viewed as chiral as they propagate in one direction only. The symmetry of time reversal, respected by Heisenberg exchange, maps $\theta_+$ and $\theta_-$ onto each other and reverses the propagation direction. A pair of chiral edge modes related by the time reversal is known as helical \cite{Hasan:2010}. 

\section{Lattice model}
\label{sec:lattice-model}

\subsection{Superexchange interactions}

\begin{figure}[th]
    \centering
    \includegraphics[width=0.95
    \columnwidth]{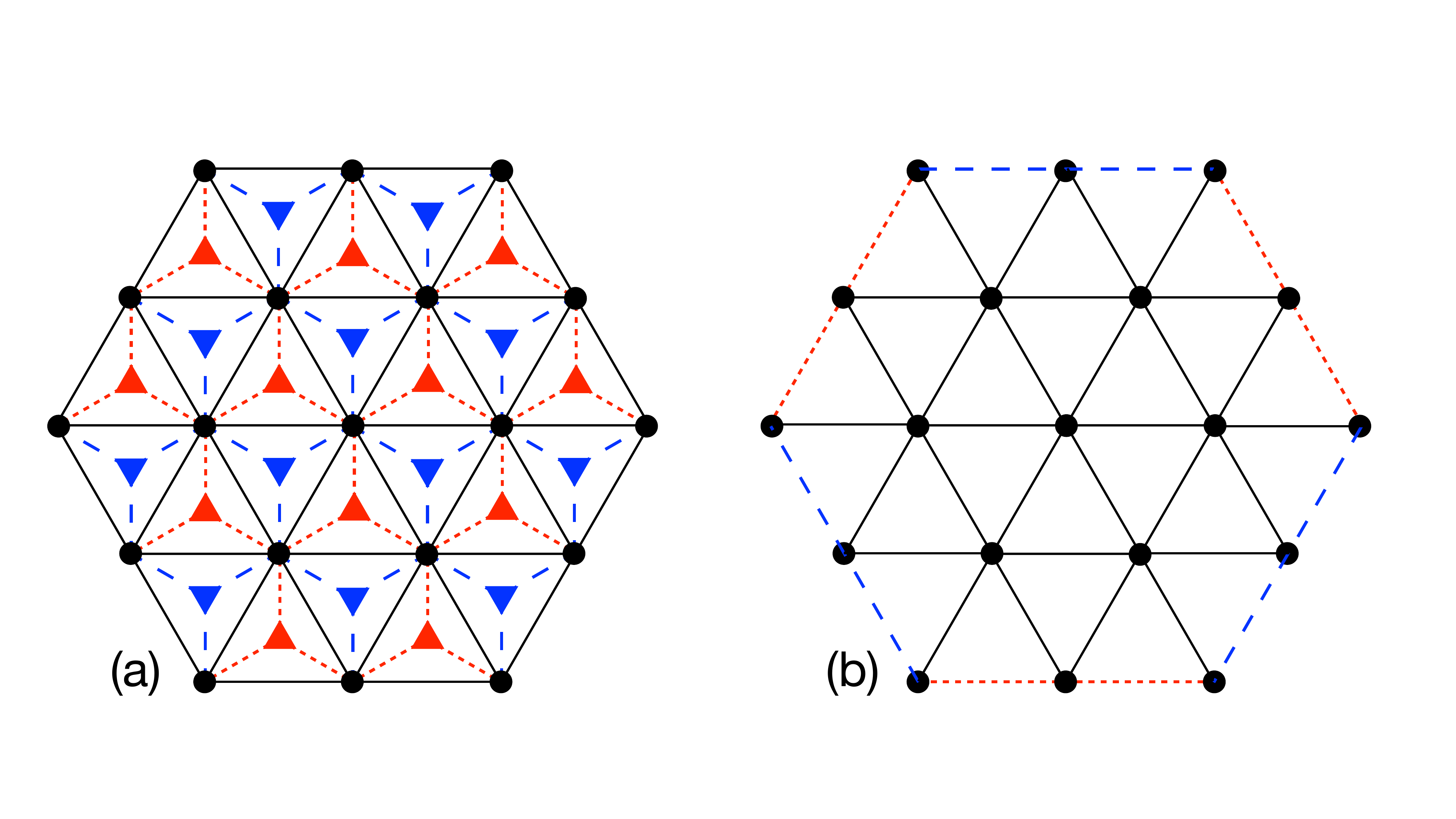}
    \caption{(a) Details of the superexchange model. Small red and blue triangles are non-magnetic ions of two different types mediate superexchange of strengths  $J^{\bigtriangleup} = J(1+\Delta)/2$ and $J^{\bigtriangledown} = J(1-\Delta)/2 $. (b) The resulting spin model has exchange interactions of strengths $J$ in the bulk (black solid lines), $J(1+\Delta)/2$ on red dotted external edges, and $J(1-\Delta)/2$ on blue dashed external edges.}  
    \label{fig:lattice-model}
\end{figure}

Heisenberg exchange interaction between two magnetic ions in a solid is often mediated by a non-magnetic ion adjacent to them, a mechanism known as superexchange. We consider a lattice model of a Heisenberg antiferromagnet with superexchange interactions illustrated in Fig.~\ref{fig:lattice-model}. Magnetic ions (black dots) reside on sites of the triangular lattice. Two different types of non-magnetic ions reside at centers of faces. Ions at the centers of $\bigtriangleup$ triangles (small red triangles) contribute superexchange of strength $J^{\bigtriangleup} = J(1+\Delta)/2$, while those on $\bigtriangledown$ triangles (small blue triangles) contribute $J^{\bigtriangledown} = J(1-\Delta)/2$. The dimensionless parameter $\Delta$ thus quantifies the difference in Heisenberg exchange for superexchange paths (red dashed and blue dotted lines) going through centers of $\bigtriangleup$ and $\bigtriangledown$ triangles. 

A pair of nearest-neighbor magnetic ions in the bulk of the system has two exchange paths connecting them, with the combined strength of exchange $J^{\bigtriangleup} + J^{\bigtriangledown} = J$ independent of $\Delta$. In contrast, two adjacent magnetic ions at the boundary are connected by just one superexchange path, so for them the net exchange strength is $J(1+\Delta)/2$ or $J(1-\Delta)/2$. We thus see that even though $\Delta$ was defined as a bulk coupling, its effect is only felt at the edge. This leads us to anticipate that the field theory of our superexchange model possesses a topological term. 

A lattice structure that could support the superexchange model is exemplified by the ternary compound BiTeI \cite{Shevelkov:1995}. It has consecutive triangular layers of Bi, Te, and I with ABC stacking. Using, say, the Bi triangular layer as a reference, we would find that Te atoms reside above the centers of its $\bigtriangleup$ triangles and I atoms below the centers of $\bigtriangledown$ triangles. Replacing Bi with a magnetic atom would realize our superexchange model. 

\subsection{Continuum approximation}

To derive the continuum approximation for the superexchange model, we begin by expressing the superexchange energy of three spins on a triangle: 

\begin{equation}
\begin{split}
u^{\bigtriangleup} &= \frac{J^{\bigtriangleup}}{2}\left(\mathbf{S}_1 + \mathbf{S}_2 + \mathbf{S}_3\right)^2, 
\\
u^{\bigtriangledown} &= \frac{J^{\bigtriangledown}}{2}\left(\mathbf{S}_1 + \mathbf{S}_2 + \mathbf{S}_3\right)^2.
\end{split}
\label{eq:superexchange-triangle}
\end{equation}
The energy of the entire lattice is obtained by summing the superexchange energies of all triangles. This form of the magnetic energy guarantees that in a ground state $\mathbf S_1 + \mathbf S_2 + \mathbf S_3 = 0$ for all triangles, including those on the boundary. 

\begin{figure}[ht]
    \centering
    \includegraphics[width=0.7
    \columnwidth]{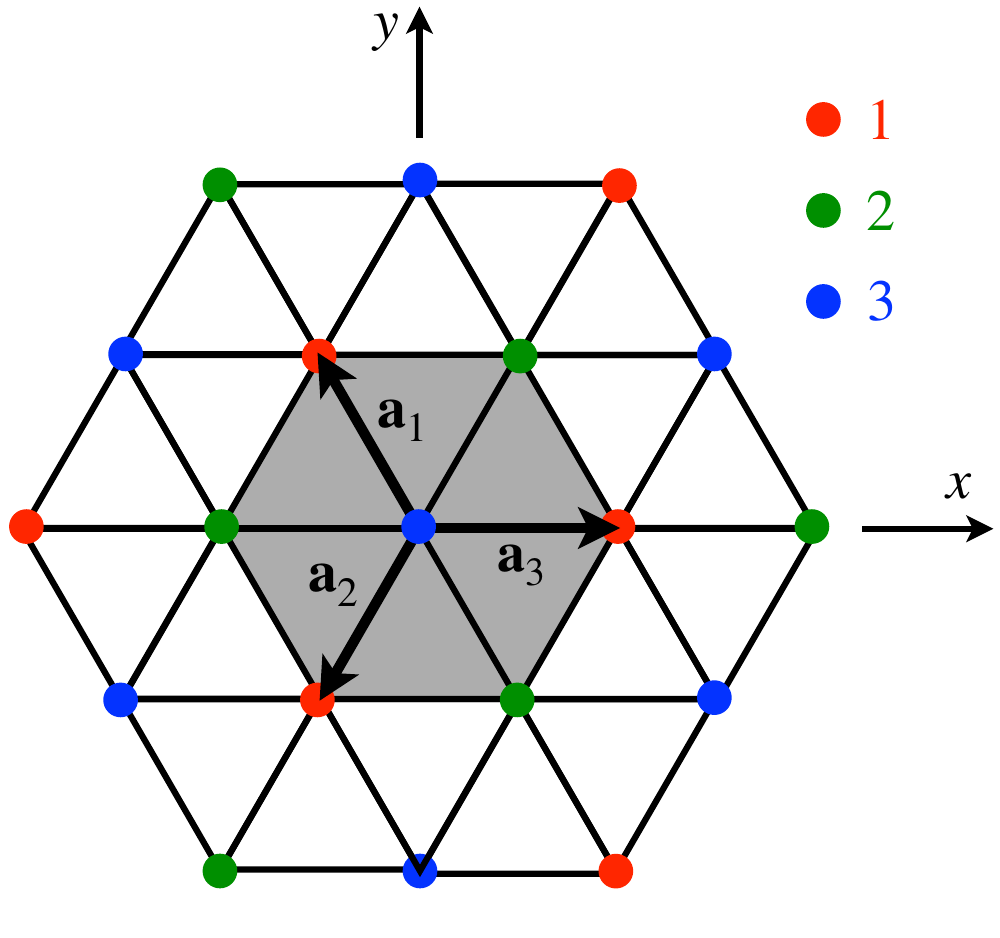}
    \caption{Geometry of the triangular lattice. Red, gren, and blue dots mark sites of magnetic sublattices 1, 2, and 3, respectively. The shaded area is the magnetic unit cell centered on a site of sublattice 3. }
    \label{fig:lattice-geometry}
\end{figure} 

A site on magnetic sublattice 3 has six nearest neighbors: three sites of sublattice 1 displaced by $\mathbf a_n$ and three sites of sublattice 2 displaced by $-\mathbf a_n$ (Fig.~\ref{fig:lattice-geometry}), where 
\begin{equation}
\mathbf a_n = a 
\left(
\cos{\frac{2\pi n}{3}}, \,
\sin{\frac{2\pi n}{3}}
\right),
\quad
n = 1, 2, 3,
\end{equation}
are primitive lattice vectors of the triangular lattice. Upon expressing the discrete spin variables $\mathbf S_n$ in terms of the sublattice magnetization fields $\mathbf m(\mathbf r)$, $\mathbf S_n = S \mathbf m_n(\mathbf r_n)$, where $\mathbf r_n$ is the position of spin $\mathbf S_n$, we obtain the superexchange energy of the $\bigtriangleup$ triangles: 
\begin{widetext}
\begin{equation} 
U^{\bigtriangleup} = 
\frac{J^{\bigtriangleup} S^2}{2}  \sum_{\mathbf{r}_3}
\left[
\mathbf{m}_1(\mathbf{r_3} + \mathbf{a}_2)
+ \mathbf{m}_2(\mathbf{r_3} - \mathbf{a}_1)
+ \mathbf{m}_3(\mathbf{r_3})
\right]^2 
+ \mbox{ cyclic permutations of } 
\{\mathbf{a}_1, \mathbf{a}_2, \mathbf{a}_3\}, 
\label{eq:up-triangle}
\end{equation}
where the sum is over sites of magnetic sublattice 3 only. We next use the gradient expansion, switch to spin-frame fields \eqref{eq:m-nx-ny-def}, and set $\mathbf m = 0$ to obtain 
\begin{equation} 
U^{\bigtriangleup} 
\approx 
\frac{9 J^\bigtriangleup S^2 a^2}{16} 
\sum_{\mathbf r_3}
(\partial_\alpha \mathbf{n}_\beta 
\cdot \partial_\alpha \mathbf{n}_\beta 
+ \partial_\alpha \mathbf{n}_\alpha 
\cdot \partial_\beta \mathbf{n}_\beta
- \partial_\beta \mathbf{n}_\alpha
\cdot \partial_\alpha \mathbf{n}_\beta). 
\end{equation}
The superexchange energy of $\bigtriangledown$ triangles is obtained along the same lines:
\begin{equation} 
U^{\bigtriangledown} 
\approx 
\frac{9 J^\bigtriangledown S^2 a^2}{16} 
\sum_{\mathbf{r_3}} 
(\partial_\alpha \mathbf{n}_\beta 
\cdot \partial_\alpha \mathbf{n}_\beta 
- \partial_\alpha \mathbf{n}_\alpha 
\cdot \partial_\beta \mathbf{n}_\beta
+\partial_\beta \mathbf{n}_\alpha \cdot \partial_\alpha \mathbf{n}_\beta). 
\end{equation}
Adding up the two contributions and passing from summation over sublattice-3 sites to integration yields the exchange energy density 
\begin{equation}
\mathcal U = 
\frac{\mu}{2} 
\partial_\alpha \mathbf{n}_\beta 
\cdot \partial_\alpha \mathbf{n}_\beta
+ \frac{\mu \Delta}{2}
(\partial_\alpha \mathbf{n}_\alpha 
\cdot \partial_\beta \mathbf{n}_\beta
- \partial_\beta \mathbf{n}_\alpha
\cdot \partial_\alpha \mathbf{n}_\beta),
\end{equation}
where $\mu = JS^2 \sqrt{3}/4$ \cite{Pradenas:2024}. This is precisely our minimal model introduced in Sec.~\ref{sec:field-theory} on heuristic grounds.  
\end{widetext}

\subsection{Spin waves in the lattice model}

We return to the lattice model with superexchange interactions to obtain its spin-wave spectrum. 

\subsubsection{Three spins on a triangle}

To begin with, we consider a single triangle. Its three spins $\mathbf S_1$, $\mathbf S_2$, and $\mathbf S_3$ are parametrized in terms of their spherical angles,
\begin{equation}
\mathbf S_n = S
(\sin{\theta_n} \cos{\phi_n}, \,
\sin{\theta_n} \sin{\phi_n}, \,
\cos{\theta_n}).
\end{equation}
Our starting point is the ground state with $\theta_n = \pi/2$ and $\phi_n = 2\pi n/3$. The spins lie in the equatorial plane and make angles of $2\pi/3$ with one another.

A weakly excited state is described by small deviation angles $\alpha$ and $\beta$:
\begin{equation}
\phi = \frac{2\pi n}{3} + \alpha, 
\quad
\theta = \frac{\pi}{2} - \beta.
\end{equation}
The Lagrangian of the system is 
\begin{equation}
L = \sum_{n=1}^3 \hbar S \cos{\theta_n} \dot{\phi}_n
- \frac{J}{2}
(\mathbf S_1 + \mathbf S_2 + \mathbf S_3)^2.
\end{equation}
Expanding it in powers of the deviation angles yields 
\begin{widetext}
\begin{equation}
L =  \hbar S 
(\dot{\alpha}_1 \beta_1 
+ \dot{\alpha}_2 \beta_2
+ \dot{\alpha}_3 \beta_3)
- \frac{JS^2}{4}
[(\alpha_2-\alpha_3)^2
    + (\alpha_3-\alpha_1)^2
    + (\alpha_1-\alpha_2)^2]
- \frac{JS^2}{2}(\beta_1 + \beta_2 + \beta_3)^2
+ \ldots
\end{equation}
\end{widetext}
up to the second order. Linearized equations of motion for $\alpha_n$ and $\beta_n$ are
\begin{equation}
\begin{split}
\hbar S \dot{\beta}_n & =
- \frac{JS^2}{2}
\sum_{m \neq n}(\alpha_n - \alpha_m),
\\
- \hbar S \dot{\alpha}_n & =
- \frac{JS^2}{2}
\sum_{m \neq n}(\beta_n + 2 \beta_m).
\end{split}
\label{eq:linear-spin-wave-eqn-three-spins}
\end{equation}

\subsubsection{Bulk spin waves}

The linear spin-wave analysis readily extends to the bulk of the sample, where exchange interaction of strength $J$ couples each spin $\mathbf S_{\mathbf r}$ to its six nearest neighbors $\mathbf S_{\mathbf r'}$. Eq.~\eqref{eq:linear-spin-wave-eqn-three-spins} generalizes to 
\begin{equation}
\begin{split}
\hbar S \dot{\beta}_{\mathbf r} & =
- \frac{JS^2}{2}
\sum_{\mathbf r'}(\alpha_{\mathbf r} - \alpha_{\mathbf r'}),
\\
- \hbar S \dot{\alpha}_{\mathbf r} & =
- \frac{JS^2}{2}
\sum_{\mathbf r'}(\beta_{\mathbf r} + 2 \beta_{\mathbf r'}).
\end{split}
\label{eq:linear-spin-wave-eqn-bulk}
\end{equation}

Using the plane-wave Ansatz, $\alpha_{\mathbf r}(t) = \alpha e^{- i \omega t + i \mathbf k \cdot \mathbf r}$ and $\beta_{\mathbf r}(t) = \beta e^{- i \omega t + i \mathbf k \cdot \mathbf r}$, yields 
\begin{equation}
\begin{split}
-i \hbar \omega \beta &= 
- JS \sum_{n=1}^3 [1 - \cos{(\mathbf k \cdot \mathbf a_n)}] \alpha,
\\
i \hbar \omega \alpha &= 
- JS \sum_{n=1}^3 [1 + 2\cos{(\mathbf k \cdot \mathbf a_n)}] \beta,
\end{split}    
\end{equation}
with lattice vectors $\mathbf a_n = 
a (\cos{\frac{2\pi n}{3}}, \, \sin{\frac{2\pi n}{3}})$. The spin-wave frequencies are then 
\begin{equation}
\left(\frac{\hbar \omega}{JS}\right)^2 
= 
\sum_{m=1}^3 [1-\cos{(\mathbf k \cdot \aaa_m)}]
\sum_{n=1}^3 [1+2\cos{(\mathbf k \cdot \aaa_n)}]
\end{equation}

\begin{figure}
    \centering
    \includegraphics[width=0.7\linewidth]{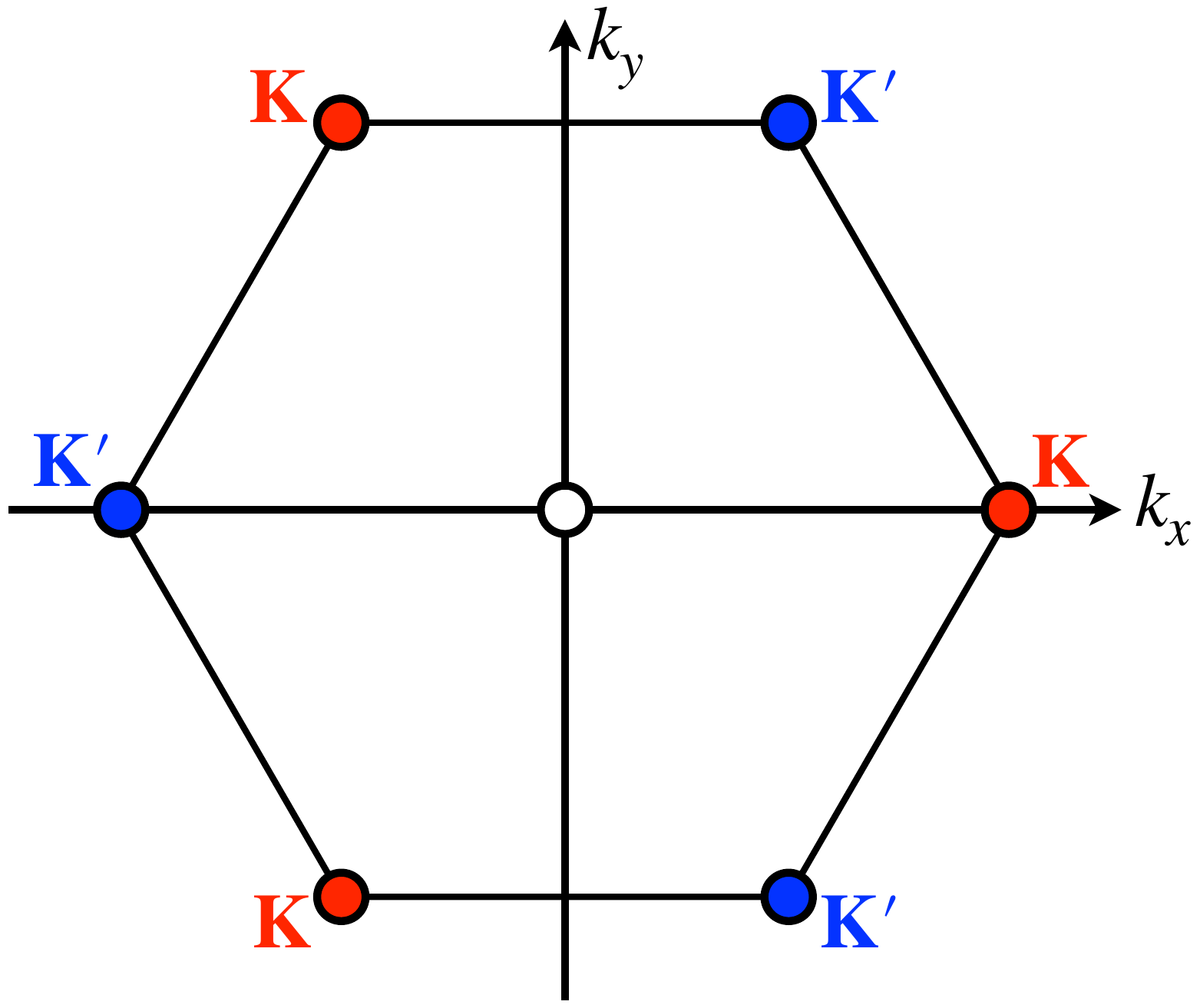}
    \caption{The frequency of spin waves vanishes at three points in the Brillouin zone: its origin 0 and corners $\mathbf K$ and $\mathbf K'$. }
    \label{fig:Brillouin-zone}
\end{figure}

The frequency vanishes at three nonequivalent points in the Brillouin zone: 
\begin{equation}
\mathbf k_\text{I} = \mathbf K, 
\quad
\mathbf k_\text{II} = \mathbf K',
\quad
\mathbf k_\text{III} = 0.
\end{equation}
Here 
\begin{equation}
\mathbf K =  
\frac{2\pi}{a}
\left(
\frac{2}{3}, \, 0
\right),
\, 
\frac{2\pi}{a}
\left(
-\frac{1}{3}, \, \pm \frac{1}{\sqrt{3}},
\right)
\label{eq:k-soft-spots}
\end{equation}
and $\mathbf K' = -\mathbf K$ are the two inequivalent corners of the Brillouin zone, Fig.~\ref{fig:Brillouin-zone}. In the vicinity of these points, the frequency rises linearly with the deviation from a soft spot, $\omega(\mathbf k + \delta \mathbf k) \sim c |\delta \mathbf k|$; the spin-wave velocities are
\begin{equation}
c_\text{I} = c_\text{II} 
= \frac{3\sqrt{3}}{2\sqrt{2}} JSa,
\quad
c_\text{III} 
= \frac{3\sqrt{3}}{2} JSa.
\label{eq:c-lattice}
\end{equation}
Spin waves in the vicinity of the soft spots \eqref{eq:k-soft-spots} can be visualized by picturing the magnetic order parameter as a rigid body [Fig.~\ref{fig:spin-frame}(b)] resembling the Mercedes-Benz logo or a thin disk defined by the coplanar spins. Spin waves near soft spot $\mathbf k_\text{III} = 0$ give rise to vibrations in the spin plane. Spin waves near soft spots $\mathbf k_\text{I}$ and $\mathbf k_\text{II}$ create a wobbling motion of the disk with the three spin arrows moving out of the spin plane with the relative phase $\phi_n = - \frac{2\pi n}{3}$ on magnetic sublattice $n$ near $\mathbf k_\text{I}$ and $\phi_n = + \frac{2\pi n}{3}$ near $\mathbf k_\text{II}$. See Supplementary Material for a video \cite{supmat}.

\subsubsection{Edge modes}

\begin{figure}
\includegraphics[width=0.95\columnwidth]{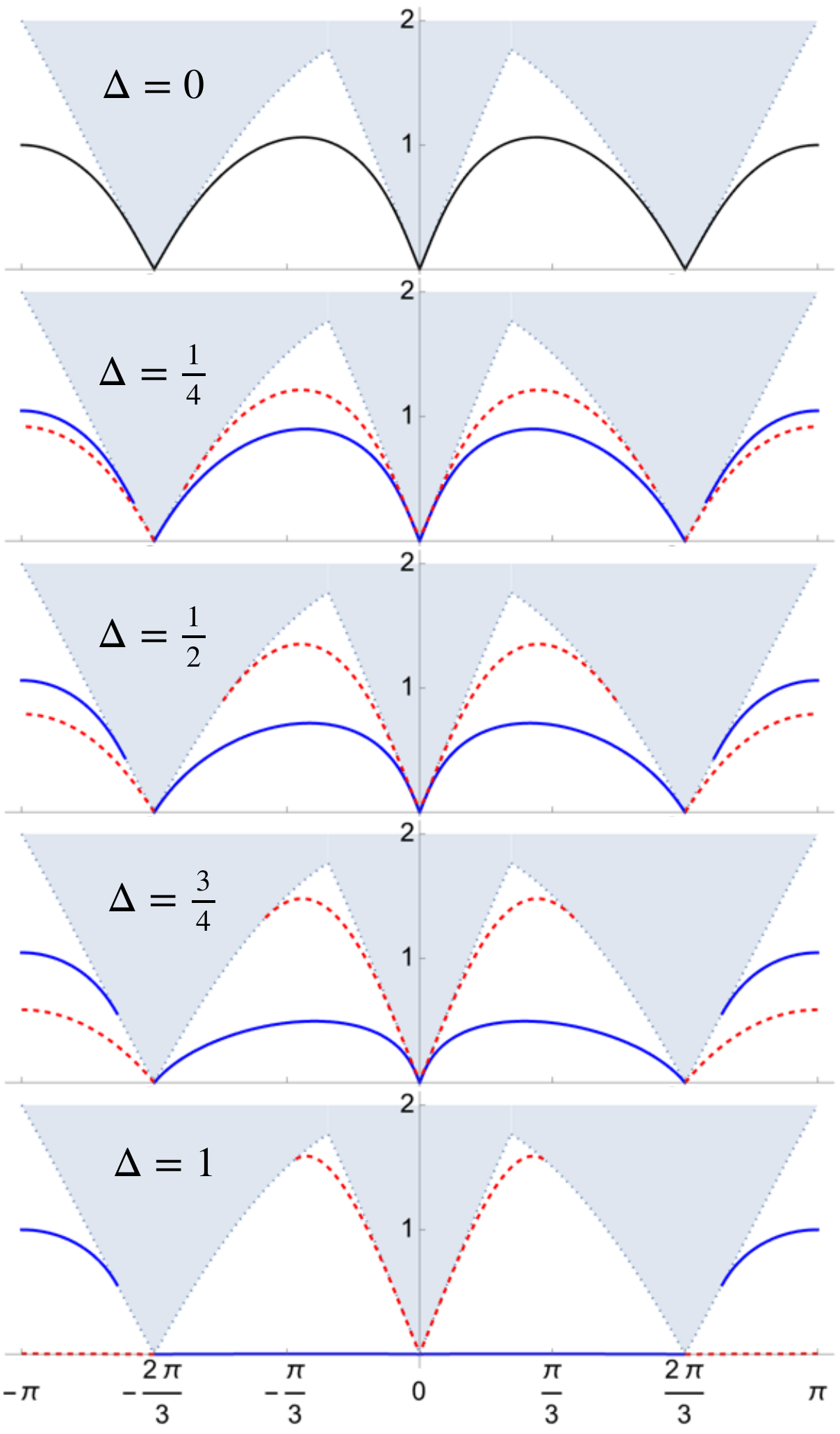}
\caption{Spin-wave spectra $\omega(k_x)$ in a strip of a triangular lattice with 200 rows. The shaded area represents bulk modes with frequencies $\omega \leq 2JS/\hbar$; the curves represent edge modes. For $\Delta = 0$, the edge modes have the same spectrum on both edges (solid black curve).   For $\Delta \neq 0$, the edge modes for the upper edge (blue solid line) and the lower edge (red dashed line) have different frequencies; in the vicinity of $k_x = \pm \frac{2\pi}{3}$, there is only one edge mode at each edge.}
\label{fig:edge-modes-strip}
\end{figure}

We have computed numerically the spectrum of linear spin waves in our lattice model for a strip with 200 rows of spins. Nearest-neighbor bonds in the bulk have exchange strength $J$. The strength of exchange for bonds on the edges is $J(1-\Delta)/2$ for the lower edge and $J(1+\Delta)/2$. Translational invariance in the direction of the strip (the $x$ direction) makes the $k_x$ component of the wavevector a good quantum number for magnons. See Appendix \ref{app:spin-waves-strip} for further details.

The frequencies $\omega(k_x)$ of bulk spin waves come down to zero at 
\begin{equation}
k_{\text{I}x} = \frac{2\pi}{3a}, 
\quad
k_{\text{II}x} = -\frac{2\pi}{3a}, 
\quad
k_{\text{III}x} = 0.
\label{eq:-kx-soft-spots}
\end{equation}
$k_x = 0$ and $\pm 2\pi/3$. In the vicinity of these points, the bulk frequencies fill cones 
\begin{equation}
\omega^2 \geq c_n^2 (k_x - k_{nx})^2 
\end{equation}
with the characteristic velocities \eqref{eq:spin-wave-velocities-triangular}.

In the absence of the topological term, $\Delta = 0$, the weakness of the bonds on the edges creates trivial edge modes for every wavenumber $k_x$ in the Brilluoin zone, Fig.~\ref{fig:edge-modes-strip}(a). In the vicinity of each soft spot \eqref{eq:-kx-soft-spots}, we find left and right-moving edge modes on both edges. 

Turning on the topological term, $\Delta \neq 0$, significantly restructures the edge modes in the vicinity of the two soft spots at $k_x = \pm \frac{2\pi}{3a}$, Fig.~\ref{fig:edge-modes-strip}: edge modes on one side of the soft spot $k_{nx}$ are pushed up the frequency axis and become bulk modes; those on the other side are pushed down and remain bound to the edge; hence the one-way propagation. No significant changes are observed near the third soft spot $k_x = 0$.

Our minimal field-theoretic model, developed in Sec.~\ref{sec:field-theory}, is incapable of describing the edge modes in the absence of the topological term. Fortunately, a minor tweak to the minimal model resolves this difficulty. 

Even in the absence of the topological term, $\Delta = 0$, bonds along the edge have a reduced strength $J/2$ of exchange interaction. This reduction of the exchange energy at the edge lowers the energy of a magnon, thereby creating an attractive potential binding it to the edge. The effect can be captured within the continuum theory by adding an energy term localized at the boundary:
\begin{equation}
U_\text{edge} = 
- \frac{\mu b}{2}\int_{\partial \Omega} dx \, 
\partial_x \mathbf n_\alpha 
\cdot \partial_x \mathbf n_\alpha, 
\label{eq:U-edge-reduced-exchange}
\end{equation}
Here we assumed that the boundary lies at $y = $ const and introduced a microscopic length scale $b>0$ of the order of the lattice constant $a$. 

With the addition of the new term, the boundary conditions \eqref{eq:spin-waves-boundary} for spin-wave fields $\theta_i$ read 
\begin{equation}
\begin{split}
\partial_y \theta_x 
- \Delta \partial_x \theta_y 
- b \, \partial_x^2 \theta_x = 0,
\\
\partial_y \theta_y 
+ \Delta \partial_x \theta_x 
- b \, \partial_x^2 \theta_y = 0,
\\
2 \partial_y \theta_z 
- 2 b \, \partial_x^2 \theta_z = 0.
\end{split}    
\end{equation}
For $\Delta = 0$, the edge-mode Ansatz \eqref{eq:edge-mode-ansatz} and the boundary condition $\partial_y\theta_i = - b k^2 \theta_i$ yield a positive inverse localization length $\kappa = bk^2$, guaranteeing an edge mode propagating in both directions for all three spin-wave modes $\theta_i$. The spectrum of these edge modes is 
\begin{equation}
\omega_n^2 = c_n^2k^2(1 - b^2 k^2),  
\end{equation}
where $n = $ I, II, III. Their velocities approaches those of the corresponding bulk spin-waves in the limit $k \to 0$. 

Turning on the topological term has no influence on mode $\theta_z$. Modes $\theta_x$ and $\theta_y$ are hybridized into complex fields $\theta_\pm$ \eqref{eq:theta-pm} with boundary conditions
\begin{equation}
\partial_y \theta_\pm 
\pm i \Delta \partial_x \theta_\pm 
- b \, \partial^2_x \theta_\pm = 0.
\end{equation}
An edge mode \eqref{eq:edge-mode-ansatz} has the inverse localization length $\kappa_\pm = bk^2 \mp \Delta k$. For one sign of $k$, namely $\mp \Delta k > 0$, the edge mode becomes even more strongly bound to the edge. For the other, $\mp \Delta k < 0$, it is loosened and even expelled entirely into the bulk at low wavenumbers, where the linear term $\mp \Delta k$ dominates over the quadratic one. In the range of wavenumbers 
\begin{equation}
- \frac{|\Delta|}{b} < k < \frac{|\Delta|}{b},    
\end{equation}
the edge mode exists for only one sign of $k$, when the inverse localization length $\kappa>0$. For $|k| \ll |\Delta|/b$, the topological term dominates over the lattice-scale energy correction \eqref{eq:U-edge-reduced-exchange}, and the field theory paints the correct picture. 

\section{Nonlinear spin waves}
\label{sec:nonlinear}

Our analysis of edge modes in the previous sections was was based on linearized equations of motion and is thus applicable to spin waves of small amplitudes. Here we shall present an exact solution for edge modes of an arbitrary amplitude. 

\subsection{Euler-angle parametrization}

The orientation of the spin frame $\{\mathbf n_x, \mathbf n_y, \mathbf n_z\}$ can be expressed in terms of Euler angles. Starting with the reference state wherein the spin frame is aligned with a set of orthonormal vectors $\{\mathbf e_x, \mathbf e_y, \mathbf e_z\}$, we rotate the frame through angle $\phi$ about $\mathbf n_z$, then through $\theta$ about $\mathbf{n}_x$, and lastly through $\psi$ about $\mathbf{n}_z$. The Euler angles $\phi$, $\theta$, and $\psi$ are our new independent fields. 

The Lagrangian of the Heisenberg antiferromagnet on a triangular lattice reads 
\begin{equation}
\begin{split}
\mathcal{L} & =
\frac{\rho}{2} 
\left[ 
\dot{\theta}^2 
+ \sin^2{\theta} \, \dot{\phi}^2  
+ (\dot{\psi}+ \cos{\theta} \, \dot{\phi})^2 
\right] 
\\
& - \frac{\mu}{2} 
\left[ 
(\nabla \theta)^2 + \sin^2{\theta}(\nabla \phi)^2  
+ 2 ( \nabla \psi +\cos{\theta}\, \nabla \phi)^2 
\right] 
\\
& - \frac{\mu\Delta}{2} 
\sin{\theta} 
(\partial_x \theta \, \partial_y \phi
- \partial_y \theta \, \partial_x\phi),
\end{split}
\label{eq:L-Euler-angles}
\end{equation}
where the last term is topological.

\subsection{Edge-wave Ansatz}

As in previous sections, we consider an antiferromagnet with a straight boundary parallel to the $x$-axis. To be more specific, we take the boundary to be the line $y = 0$, with the magnetic medium in the semi-plane $y>0$. 

Angle $\theta$ describes the deviation of the spin-frame axis $\mathbf n_z$ from its ground-state direction $\mathbf e_z$ and thus serves as the amplitude of the edge mode. We assume that this amplitude depends on the distance $y$ from the edge, hence $\theta = \theta(y)$. The other two angles $\phi$ and $\psi$ paramertize oscillations of the spin frame in space and time. For a wave propagating aling the edge, $\phi = \phi(t,x)$ and $\psi = \psi(t,x)$. 

Under these assumptions, the equations of motion for fields $\psi$ and $\phi$ read, respectively,
\begin{equation}
\begin{split}
\rho(\ddot{\psi} + \cos{\theta} \, \ddot{\phi})
&= 2\mu \psi'' + 2\mu \cos{\theta} \, \phi'',
\\
\rho( \cos{\theta} \, \ddot{\psi} + \ddot{\phi})
&= 2 \mu \cos{\theta} \, \psi''
+ \mu (1+\cos^2{\theta})\phi'' .
\end{split}    
\end{equation}
The dot and prime stand for $\partial_t$ and $\partial_x$, respectively. A simple class of solutions $\psi(t,x)$ and $\phi(t,x)$ are linear functions of $t$ and $x$. 

A further restriction on $\psi$ and $\phi$ comes from the asymptotic behavior of an edge mode. Far from the boundary, its amplitude decays, 
\begin{equation}
\theta(y) \rightarrow 0 
\mbox{ as }
y \to +\infty.
\label{eq:theta-to-0-at-infty}
\end{equation}
The orientation of the spin frame is then parametrized by a single rotation through angle $\phi+\psi$ about $\mathbf n_z = \mathbf e_z$. In order to the system to be in a uniform ground state far from the boundary, we must demand that $\phi + \psi = 0$. This yields the edge-mode Ansatz,
\begin{equation}
\phi = - \psi = kx - \omega t, 
\quad
\theta = \theta(y).
\end{equation}
It describes a spin frame tilted by angle $\theta(y)$ about the rotating axis 
\begin{equation}
\mathbf e(t) = 
\mathbf e_x \cos{\omega t}
- \mathbf e_y \sin{\omega t}.
\end{equation}
For $\omega>0$, the unit vector $\mathbf e(t)$ rotates clockwise. 

\subsection{Boundary conditions}

The first variation of the action contains a boundary term: 
\begin{equation}
\delta S_\text{edge} 
= \int dt \, dx \, \mu 
[\theta'(y) - \Delta k \sin{\theta(y)}] \delta \theta(y) 
\big|_{y=0}.
\end{equation}
Setting this variation to zero yields the boundary condition
\begin{equation}
\theta' - \Delta k \sin{\theta} = 0 
\mbox{ at } y = 0.
\label{eq:theta-boundary-condition}
\end{equation}

\subsection{Profile of the edge mode}

The equation of motion for the amplitude profile $\theta(y)$ of the edge mode is 
\begin{equation}
\mu \theta''
+ \rho \omega^2 \sin{\theta} 
+ \mu k^2 \sin{\theta} \cos{\theta} 
- 2 \mu k^2 \sin{\theta}
= 0.
\end{equation}
It has a first integral, which can be obtained directly from the Lagrangian \eqref{eq:L-Euler-angles} combined with the edge-mode Ansatz \eqref{eq:edge-mode-ansatz}: 
\begin{equation}
\mathcal E = \frac{\mu {\theta'}^2}{2} 
+ \frac{\rho \omega^2 - \mu k^2}{2} \sin^2{\theta}
+ \frac{\rho \omega^2 - 2\mu k^2}{2} (1-\cos{\theta})^2.
\end{equation}
The asymptotic behavior \eqref{eq:theta-to-0-at-infty} implies that $\mathcal E = 0$. Combining this result with the boundary condition \eqref{eq:theta-boundary-condition} yields the spectrum of the edge mode, 
\begin{equation}
\begin{split}
\omega^2 &= c_e^2 k^2,   
\\
c_e^2 &= 
c_\text{I}^2 
[1-\Delta^2 + (1+\Delta^2)\sin^2{(\Theta/2)}].
\end{split}
\label{eq:edge-mode-speed-finite}
\end{equation}
Here $c_\text{I}$ is the speed of bulk spin waves \eqref{eq:spin-wave-velocities-triangular} and $\Theta \equiv \theta(0)$ is the amplitude at the edge. In the limit of a small amplitude, $\Theta \to 0$, we recover the previously obtained result \eqref{eq:edge-mode-speed-infinitesimal} of the linear spin-wave theory.

Far from the edge, the amplitude $\theta(y)$ is small. Expanding the condition $\mathcal E = 0$ to the lowest nonvanishing order in $\theta$, we find ${\theta'}^2 \sim \kappa^2 \theta^2$, which gives $\theta(y) \sim C e^{\pm \kappa y}$ for $y \to +\infty$. The inverse localization length $\kappa$ is given by the equation 
\begin{equation}
\kappa^2 = k^2
[\Delta^2 \cos^2{(\Theta/2)} - \sin^2{(\Theta/2)}].
\end{equation}
In the limit $\Theta \to 0$, this result agrees with Eq.~\eqref{eq:kappa-Delta-k-linear} of the linear spin-wave theory. 

The edge amplitude $\Theta$ has a threshold beyond which the edge mode is expelled into the bulk. At that point, the localization length diverges, which means $\kappa = 0$. The threshold amplitude is determined by the condition 
\begin{equation}
\tan^2{(\Theta/2)} = \Delta^2.    
\end{equation}
At that amplitude, the speed of the edge mode becomes equal to that of the bulk mode, $c_e^2 = c_\text{I}^2$. 

\section{Discussion}
\label{sec:discussion}

We have presented a theoretical study of helical edge modes in a Heisenberg antiferromagnet on a triangular lattice. These modes propagate in one direction along the edge for clockwise circular polarization and in the opposite direction for counterclockwise polarization. These edge modes do not require interactions that break the global symmetry of spin rotations, such as the Dzyaloshinskii--Moriya term, and do not depend on nontrivial band topology of magnons. 

A first clue for the existence of these edge modes came from the continuum theory of the Heisenberg antiferromagnet on a triangular lattice. The exchange energy, quadratic in the gradient of staggered magnetization fields, allows for a term that is silent in the bulk and can be expressed as an edge term or as a topological invariant---the skyrmion number of the vector chirality. The edge term modifies the boundary conditions for spin waves, creating edge modes that propagate in one direction along the boundary for each circular polarization. The time reversal symmetry of Heisenberg exchange reverses both the polarization and direction of propagation. This pairing of counterpropagating modes is reminiscent of topological insulators in two dimensions, where electrons with opposite spin travel in opposite directions along the edge. 

Building on this field-theoretic argument, we have found a lattice model with Heisenberg superexchange interactions whose continuum description indeed has precisely this topological term. In this model, superexchange is mediated by two different types of non-magnetic ions residing at the centers of triangles of two different orientations: $J(1+\Delta)/2$ on triangles of the orientation $\bigtriangleup$ and 
$J(1-\Delta)/2$ on those with the orientation $\bigtriangledown$. The topological term is proportional to the dimensionless disparity parameter $\Delta$. Although the disparity does not manifest itself in the bulk, where each pair of magnetic ions has two superexchange paths adding up to a total exchange integral of $J$, it becomes effective at the edge, where only one superexchange path is available. We have studied the spectrum of linear spin waves in this lattice model numerically and confirmed the existence of helical edge modes. All the salient features of the spin-wave spectra are explained within the framework of a continuum model. 

It is instructive to trace the formation of the helical edge modes in the extreme disparity limit, $\Delta \to 1$. At $\Delta = 1$, the lattice model in an infinite strip with straight edges has zero modes for each wavenumber $k_x$ along the direction of the strip---see Appendix \ref{app:zero-modes-Delta=1}. The zero modes are localized at the upper edge for $|k_x| < \frac{2\pi}{3a}$ and at the lower edge for $\frac{2\pi}{3a} < |k_x| \leq \frac{\pi}{a}$, see the bottom panel of Fig.~\ref{fig:edge-modes-strip}. Once $\Delta$ is reduced below 1, these formerly zero modes are lifted to finite frequencies away from the soft spots $k_x = \pm \frac{2\pi}{3a}$, see the spin-wave spectra for $\Delta = 3/4$ in Fig.~\ref{fig:edge-modes-strip}. The edge modes on the right side of the soft spot $k_x = \frac{2\pi}{3a}$ propagate in the $+x$ direction along the lower edge, whereas those on the left side of the same soft spot propagate in the $-x$ direction along the upper edge. The edge modes near the other soft spot, $k_x = - \frac{2\pi}{3a}$, are time-reversed copies of these.

The theory presented in this paper leaves out a couple of important issues that we plan to address in future publications. The first of these is the nature of the quantum observable that distinguishes magnons moving in opposite directions along the edge. At the classical level, they are spin waves with two opposite circular polarizations, as explained in Sec.~\ref{sec:field-theory:helical-modes}. This question applies to both edge and bulk magnons: two out of three spin-wave branches have the same velocity, as seen in Eq.~\eqref{eq:spin-wave-velocities-triangular}. This magnon degeneracy hints at the presence of a symmetry whose generator can serve as the missing quantum number. In a Heisenberg ferromagnet, or a collinear antiferromagnet, the magnetic order parameter spontaneously breaks the global $SO(3)$ symmetry of global spin rotations down to the $SO(2)$ symmetry of rotations about the direction of the ordered spins, say the $z$-axis. This residual symmetry yields a quantum number for magnons, the magnon spin $S_z$, which equals $-1$ for a Heisenberg ferromagnet and $\pm 1$ for a collinear Heisenberg antiferromagnet. In a non-collinear antiferromagnet, such as the Heisenberg antiferromagnet on a triangular lattice, the magnetic order breaks the global $SO(3)$ symmetry of spin rotations completely: there is no axis of rotation that would leave the coplanar magnetic order invariant. It may thus appear that no residual symmetry exists to define the missing quantum number. The paradox may be resolved through a careful analysis of the rotational symmetry, which combines rotations with respect to a global fixed frame, $\{\mathbf e_x, \mathbf e_y, \mathbf e_z\}$, with rotations with respect to the spin frame, $\{\mathbf n_x, \mathbf n_y, \mathbf n_z\}$. The generators of these two sets of rotations are projections of the spin onto the global and spin-frame axes, respectively. 

Another important theme is the stability of the helical edge modes. In the idealized setting considered in this paper---a perfectly straight sample edge---the edge modes are protected by the symmetry of translations along the edge. However, if the edge is rough, then momentum conservation is lost. Elastic scattering from edge imperfections would then hybridize edge modes with a continuum of bulk modes of the same frequency (see the spectrum in Fig.~\ref{fig:strip-spectrum-minimal-model}). Thus edge roughness would lead to a finite lifetime of the helical edge modes. An estimate of the lifetime will be published separately. 

\section*{Acknowledgments} We thank Hua Chen, Boris Ivanov, and Se Kwon Kim for useful discussions. The research has been supported by the U.S. Department of Energy under Award No. DE-SC0019331 and by the U. S. National Science Foundation under Grant No. NSF PHY-2309135.

\appendix

\begin{widetext}

\section{Spin waves in an infinite strip}
\label{app:spin-waves-strip}

We consider a lattice strip infinite along the $x$ direction and containing $\ell$ rows in the $y$ direction. We assume that the $x$ dependence of both $\alpha$ and $\beta$ is in the form $e^{i k_x x}$, where $x$ is the coordinate of the lattice site. Then amplitudes are labeled by the row index $n$. Equations of motion (\ref{eq:linear-spin-wave-eqn-bulk}) read 
\begin{equation}
\begin{split}
- \dot{\beta}_n &= 
\left(
    \alpha_n - \alpha_{n-1} \cos{\textstyle{\frac{k_x}{2}}}
\right)
+ \left(
    \alpha_n - \alpha_n \cos{k_x}
\right)
+ \left(
    \alpha_n - \alpha_{n+1} \cos{\textstyle{\frac{k_x}{2}}}
\right),
\\
\dot{\alpha}_n &= 
\left(
    \beta_n + 2\beta_{n-1} \cos{\textstyle{\frac{k_x}{2}}}
\right)
+ \left(
    \beta_n + 2\beta_n \cos{k_x}
\right)
+ \left(
    \beta_n + 2\beta_{n+1} \cos{\textstyle{\frac{k_x}{2}}}
\right).
\end{split}  
\label{eq:spin-waves-strip-bulk-1}
\end{equation}
Here we have switched to the natural units of time $\tau = \hbar/JS$ and length $a$ and grouped terms on the right-hand side by the row. 

Eqs.~(\ref{eq:spin-waves-strip-bulk-1}) apply to bulk rows $1 < n < \ell$. At the edges, $n=1$ and $\ell$, the equations are modified to reflect the absence of rows $n=0$ and $\ell+1$ and to include the modified exchange couplings along the edges, of strengths $J(1\pm\Delta)/2$: 
\begin{equation}
\begin{split}
- \dot{\beta}_1 &= 
0\left(
    \alpha_1 - \alpha_0 \cos{\textstyle{\frac{k_x}{2}}}
\right)
+ \frac{1+\Delta}{2}\left(
    \alpha_1 - \alpha_1 \cos{k_x}
\right)
+ \left(
    \alpha_1 - \alpha_{2} \cos{\textstyle{\frac{k_x}{2}}}
\right), 
\\
\dot{\alpha}_1 &= 
0 \left(
    \beta_1 + 2\beta_0 \cos{\textstyle{\frac{k_x}{2}}}
\right)
+ \frac{1+\Delta}{2}\left(
    \beta_1 + 2\beta_1 \cos{k_x}
\right)
+ \left(
    \beta_1 + 2\beta_2 \cos{\textstyle{\frac{k_x}{2}}}
\right),
\\
- \dot{\beta}_\ell &= 
\left(
    \alpha_{\ell} - \alpha_{\ell-1} \cos{\textstyle{\frac{k_x}{2}}}
\right)
+ \frac{1-\Delta}{2}\left(
    \alpha_\ell - \alpha_\ell \cos{k_x}
\right)
+ 0\left(
    \alpha_\ell - \alpha_{\ell+1} \cos{\textstyle{\frac{k_x}{2}}}
\right),
\\
\dot{\alpha}_\ell &= 
\left(
    \beta_\ell + 2\beta_{\ell-1} \cos{\textstyle{\frac{k_x}{2}}}
\right)
+ \frac{1-\Delta}{2}\left(
    \beta_\ell + 2\beta_\ell \cos{k_x}
\right)
+ 0\left(
    \beta_\ell + 2\beta_{\ell+1} \cos{\textstyle{\frac{k_x}{2}}}
\right).
\end{split}
\label{eq:spin-waves-strip-edges-1}
\end{equation}

An alternative way to obtain the dynamics of spin waves in a strip is to consider an infinite plane, thus allowing rows with $n<1$ and $n>\ell$, and impose the boundary conditions 
\begin{equation}
\begin{split}
0 &= 
\left(
    \alpha_1 - \alpha_0 \cos{\textstyle{\frac{k_x}{2}}}
\right)
+ \frac{1-\Delta}{2}\left(
    \alpha_1 - \alpha_1 \cos{k_x a}
\right), 
\\
0 &= 
\left(
    \beta_1 + 2\beta_0 \cos{\textstyle{\frac{k_x}{2}}}
\right)
+ \frac{1-\Delta}{2}\left(
    \beta_1 + 2\beta_1 \cos{k_x a}
\right),
\\
0 &= 
\frac{1+\Delta}{2}\left(
    \alpha_\ell - \alpha_\ell \cos{k_x}
\right)
+ \left(
    \alpha_\ell - \alpha_{\ell+1} \cos{\textstyle{\frac{k_x}{2}}}
\right),
\\
0 &= 
\frac{1+\Delta}{2}\left(
    \beta_\ell + 2\beta_\ell \cos{k_x}
\right)
+ \left(
    \beta_\ell + 2\beta_{\ell+1} \cos{\textstyle{\frac{k_x}{2}}}
\right).
\end{split} 
\label{eq:boundary-conditions}
\end{equation}
The reader can check that combining the bulk equations of motion \eqref{eq:spin-waves-strip-bulk-1} for $n = 1$ or $\ell$ with boundary conditions \eqref{eq:boundary-conditions} indeed yields the equations of motion for the respective edges \eqref{eq:spin-waves-strip-edges-1}.

Assuming the time dependence $\alpha_n(t) = \alpha_n \sin{\omega t}$ and $\beta_n(t) = \beta_n \cos{\omega t}$ yields linear algebraic equations for the bulk,
\begin{equation}
\begin{split}
- \cos{\textstyle{\frac{k_x}{2}}} \, \alpha_{n-1}
+ (3 - \cos{k_x}) \, \alpha_n 
- \cos{\textstyle{\frac{k_x}{2}}} \, \alpha_{n+1}
& = \omega \beta_n,
\\
2 \cos{\textstyle{\frac{k_x}{2}}} \, \beta_{n-1}
+ (3 + 2 \cos{k_x}) \, \beta_n 
+2 \cos{\textstyle{\frac{k_x}{2}}} \, \beta_{n+1}
& = \omega \alpha_n,
\end{split}    
\label{eq:spin-waves-strip-bulk-2}
\end{equation}
and for the edges,
\begin{equation}
\begin{split}
\left(
1+\frac{1+\Delta}{2}(1 - \cos{k_x})
\right) \alpha_1 
- \cos{\textstyle{\frac{k_x}{2}}} \, \alpha_2
& = \omega \beta_1,
\\
\left(
1 + \frac{1+\Delta}{2}(1 + 2\cos{k_x}) 
\right) \beta_1 
+ 2 \cos{\textstyle{\frac{k_x}{2}}} \, \beta_2
& = \omega \alpha_1,
\\
- \cos{\textstyle{\frac{k_x}{2}}} \, \alpha_{\ell-1}
+ \left(
1-\frac{1-\Delta}{2}(1 - \cos{k_x})
\right) \alpha_\ell
& = \omega \beta_\ell,
\\
2 \cos{\textstyle{\frac{k_x}{2}}} \, \beta_{\ell-1}
+ \left(
1 + \frac{1-\Delta}{2}(1 + 2\cos{k_x}) 
\right) \beta_\ell
& = \omega \alpha_\ell.
\end{split}
\end{equation}

Combining the amplitudes $\alpha_n$ and $\beta_n$ into a vector $\psi = (\alpha_1, \beta_1, \ldots, \alpha_\ell, \beta_\ell)$, we obtain a generalized eigenvalue problem 
\begin{equation}
H \psi = \omega K \psi,    
\end{equation}
where $H$ is a sparse symmetric matrix with nonzero entries in the bulk ($1 < n < \ell$),
\begin{equation}
\begin{split}
H_{2n-3,2n-1} =  H_{2n-1,2n-3}
= H_{2n-1,2n+1} =  H_{2n+1,2n-1}
& = - \cos{\textstyle{\frac{k_x}{2}}},   
\\
H_{2n-1,2n-1} & = 3 - \cos{k_x},
\\
H_{2n-2,2n} =  H_{2n,2n-2}
= H_{2n,2n+2} =  H_{2n+2,2n}
& = 2 \cos{\textstyle{\frac{k_x}{2}}},
\\
H_{2n,2n} & = 3 + 2 \cos{k_x},
\end{split}
\end{equation}
and at the edges,
\begin{equation}
\begin{split}
H_{1,1} 
& = 1+\frac{1+\Delta}{2}(1 - \cos{k_x}),
\\
H_{2,2} 
& = 1+\frac{1+\Delta}{2}(1 + 2\cos{k_x}),
\\
H_{2\ell-1,2\ell-1} 
& = 1+\frac{1-\Delta}{2}(1 - \cos{k_x}),
\\
H_{2\ell,2\ell} 
& = 1+\frac{1-\Delta}{2}(1 + 2\cos{k_x}).
\end{split}    
\end{equation}
Sparse symmetric matrix $K$ has nonzero matrix elements 
\begin{equation}
K_{2n-1,2n} = K_{2n,2n-1} = 1, 
\quad
1 \leq n \leq \ell.
\end{equation}

\section{Zero modes on the brink of instability}
\label{app:zero-modes-Delta=1}

When the topological term strength $\Delta$ reaches 1, the velocity of edge modes \eqref{eq:edge-mode-speed-infinitesimal} vanishes, signaling an instability of magnetic order. The numerically obtained spectrum of spin waves in an infinite strip (Fig.~\ref{fig:edge-modes-strip}) shows edge waves with $\omega = 0$ for the entire one-dimensional Brilloin zone at $\Delta = 1$. We shall derive this numerical result analytically. 

Consider equations for spin waves in the bulk \eqref{eq:spin-waves-strip-bulk-2}. For $\omega = 0$, equations for the $\alpha$ and $\beta$ amplitudes decouple. In what follows, we focus on the $\beta$ modes representing spin oscillations normal to the spin plane. 

A static wave that oscillates in the $x$ direction at the wavenumber $k_x$ and varies exponentially in the $y$ direction, $\beta_n = \beta_0 q^n$, the bulk equation of motion \eqref{eq:spin-waves-strip-bulk-2} reads 
\begin{equation}
2 \cos{\textstyle{\frac{k_x}{2}}}
(q+q^{-1}) + (3+2\cos{k_x}) = 0.
\end{equation}
It has two solutions,
\begin{equation}
q = - 2 \cos{\textstyle{\frac{k_x}{2}}},
\quad
q = - \frac{1}{2 \cos{\textstyle{\frac{k_x}{2}}}}.
\end{equation}
At $\Delta = 1$, the boundary conditions \eqref{eq:boundary-conditions} for the $\beta$ modes are compatible with one of these solutions, $q = - 2 \cos{\textstyle{\frac{k_x}{2}}}$. For $|k_x| < \frac{2\pi}{3}$, the quotient $|q|>1$, so the zero mode is localized at the upper edge. For $|k_x| > \frac{2\pi}{3}$, $|q|<1$, and the zero mode is localized at the lower edge, in agreement with the numerical results, Fig.~\ref{fig:edge-modes-strip}.

\end{widetext}

\bibliography{references}

\begin{thebibliography}{13}%
\makeatletter
\providecommand \@ifxundefined [1]{%
 \@ifx{#1\undefined}
}%
\providecommand \@ifnum [1]{%
 \ifnum #1\expandafter \@firstoftwo
 \else \expandafter \@secondoftwo
 \fi
}%
\providecommand \@ifx [1]{%
 \ifx #1\expandafter \@firstoftwo
 \else \expandafter \@secondoftwo
 \fi
}%
\providecommand \natexlab [1]{#1}%
\providecommand \enquote  [1]{``#1''}%
\providecommand \bibnamefont  [1]{#1}%
\providecommand \bibfnamefont [1]{#1}%
\providecommand \citenamefont [1]{#1}%
\providecommand \href@noop [0]{\@secondoftwo}%
\providecommand \href [0]{\begingroup \@sanitize@url \@href}%
\providecommand \@href[1]{\@@startlink{#1}\@@href}%
\providecommand \@@href[1]{\endgroup#1\@@endlink}%
\providecommand \@sanitize@url [0]{\catcode `\\12\catcode `\$12\catcode `\&12\catcode `\#12\catcode `\^12\catcode `\_12\catcode `\%12\relax}%
\providecommand \@@startlink[1]{}%
\providecommand \@@endlink[0]{}%
\providecommand \url  [0]{\begingroup\@sanitize@url \@url }%
\providecommand \@url [1]{\endgroup\@href {#1}{\urlprefix }}%
\providecommand \urlprefix  [0]{URL }%
\providecommand \Eprint [0]{\href }%
\providecommand \doibase [0]{https://doi.org/}%
\providecommand \selectlanguage [0]{\@gobble}%
\providecommand \bibinfo  [0]{\@secondoftwo}%
\providecommand \bibfield  [0]{\@secondoftwo}%
\providecommand \translation [1]{[#1]}%
\providecommand \BibitemOpen [0]{}%
\providecommand \bibitemStop [0]{}%
\providecommand \bibitemNoStop [0]{.\EOS\space}%
\providecommand \EOS [0]{\spacefactor3000\relax}%
\providecommand \BibitemShut  [1]{\csname bibitem#1\endcsname}%
\let\auto@bib@innerbib\@empty
\bibitem [{\citenamefont {Halperin}(1982)}]{Halperin:1982}%
  \BibitemOpen
  \bibfield  {author} {\bibinfo {author} {\bibfnamefont {B.~I.}\ \bibnamefont {Halperin}},\ }\bibfield  {title} {\bibinfo {title} {Quantized {Hall} conductance, current-carrying edge states, and the existence of extended states in a two-dimensional disordered potential},\ }\href {https://doi.org/10.1103/PhysRevB.25.2185} {\bibfield  {journal} {\bibinfo  {journal} {Phys. Rev. B}\ }\textbf {\bibinfo {volume} {25}},\ \bibinfo {pages} {2185} (\bibinfo {year} {1982})}\BibitemShut {NoStop}%
\bibitem [{\citenamefont {Hasan}\ and\ \citenamefont {Kane}(2010)}]{Hasan:2010}%
  \BibitemOpen
  \bibfield  {author} {\bibinfo {author} {\bibfnamefont {M.~Z.}\ \bibnamefont {Hasan}}\ and\ \bibinfo {author} {\bibfnamefont {C.~L.}\ \bibnamefont {Kane}},\ }\bibfield  {title} {\bibinfo {title} {Colloquium: Topological insulators},\ }\href {https://doi.org/10.1103/RevModPhys.82.3045} {\bibfield  {journal} {\bibinfo  {journal} {Rev. Mod. Phys.}\ }\textbf {\bibinfo {volume} {82}},\ \bibinfo {pages} {3045} (\bibinfo {year} {2010})}\BibitemShut {NoStop}%
\bibitem [{\citenamefont {Haldane}\ and\ \citenamefont {Raghu}(2008)}]{Haldane:2008}%
  \BibitemOpen
  \bibfield  {author} {\bibinfo {author} {\bibfnamefont {F.~D.~M.}\ \bibnamefont {Haldane}}\ and\ \bibinfo {author} {\bibfnamefont {S.}~\bibnamefont {Raghu}},\ }\bibfield  {title} {\bibinfo {title} {Possible realization of directional optical waveguides in photonic crystals with broken time-reversal symmetry},\ }\href {https://doi.org/10.1103/PhysRevLett.100.013904} {\bibfield  {journal} {\bibinfo  {journal} {Phys. Rev. Lett.}\ }\textbf {\bibinfo {volume} {100}},\ \bibinfo {pages} {013904} (\bibinfo {year} {2008})}\BibitemShut {NoStop}%
\bibitem [{\citenamefont {Wang}\ \emph {et~al.}(2009)\citenamefont {Wang}, \citenamefont {Chong}, \citenamefont {Joannopoulos},\ and\ \citenamefont {Solja{\v{c}}i{\'{c}}}}]{Wang:2009}%
  \BibitemOpen
  \bibfield  {author} {\bibinfo {author} {\bibfnamefont {Z.}~\bibnamefont {Wang}}, \bibinfo {author} {\bibfnamefont {Y.}~\bibnamefont {Chong}}, \bibinfo {author} {\bibfnamefont {J.~D.}\ \bibnamefont {Joannopoulos}},\ and\ \bibinfo {author} {\bibfnamefont {M.}~\bibnamefont {Solja{\v{c}}i{\'{c}}}},\ }\bibfield  {title} {\bibinfo {title} {Observation of unidirectional backscattering-immune topological electromagnetic states},\ }\href {https://doi.org/10.1038/nature08293} {\bibfield  {journal} {\bibinfo  {journal} {Nature}\ }\textbf {\bibinfo {volume} {461}},\ \bibinfo {pages} {772} (\bibinfo {year} {2009})}\BibitemShut {NoStop}%
\bibitem [{\citenamefont {Matsumoto}\ and\ \citenamefont {Murakami}(2011)}]{Matsumoto:2011}%
  \BibitemOpen
  \bibfield  {author} {\bibinfo {author} {\bibfnamefont {R.}~\bibnamefont {Matsumoto}}\ and\ \bibinfo {author} {\bibfnamefont {S.}~\bibnamefont {Murakami}},\ }\bibfield  {title} {\bibinfo {title} {Theoretical prediction of a rotating magnon wave packet in ferromagnets},\ }\href {https://doi.org/10.1103/PhysRevLett.106.197202} {\bibfield  {journal} {\bibinfo  {journal} {Phys. Rev. Lett.}\ }\textbf {\bibinfo {volume} {106}},\ \bibinfo {pages} {197202} (\bibinfo {year} {2011})}\BibitemShut {NoStop}%
\bibitem [{\citenamefont {Shindou}\ \emph {et~al.}(2013)\citenamefont {Shindou}, \citenamefont {Matsumoto}, \citenamefont {Murakami},\ and\ \citenamefont {Ohe}}]{Shindou:2013}%
  \BibitemOpen
  \bibfield  {author} {\bibinfo {author} {\bibfnamefont {R.}~\bibnamefont {Shindou}}, \bibinfo {author} {\bibfnamefont {R.}~\bibnamefont {Matsumoto}}, \bibinfo {author} {\bibfnamefont {S.}~\bibnamefont {Murakami}},\ and\ \bibinfo {author} {\bibfnamefont {J.-i.}\ \bibnamefont {Ohe}},\ }\bibfield  {title} {\bibinfo {title} {Topological chiral magnonic edge mode in a magnonic crystal},\ }\href {https://doi.org/10.1103/PhysRevB.87.174427} {\bibfield  {journal} {\bibinfo  {journal} {Phys. Rev. B}\ }\textbf {\bibinfo {volume} {87}},\ \bibinfo {pages} {174427} (\bibinfo {year} {2013})}\BibitemShut {NoStop}%
\bibitem [{\citenamefont {Zhang}\ \emph {et~al.}(2013)\citenamefont {Zhang}, \citenamefont {Ren}, \citenamefont {Wang},\ and\ \citenamefont {Li}}]{Zhang:2013}%
  \BibitemOpen
  \bibfield  {author} {\bibinfo {author} {\bibfnamefont {L.}~\bibnamefont {Zhang}}, \bibinfo {author} {\bibfnamefont {J.}~\bibnamefont {Ren}}, \bibinfo {author} {\bibfnamefont {J.-S.}\ \bibnamefont {Wang}},\ and\ \bibinfo {author} {\bibfnamefont {B.}~\bibnamefont {Li}},\ }\bibfield  {title} {\bibinfo {title} {Topological magnon insulator in insulating ferromagnet},\ }\href {https://doi.org/10.1103/PhysRevB.87.144101} {\bibfield  {journal} {\bibinfo  {journal} {Phys. Rev. B}\ }\textbf {\bibinfo {volume} {87}},\ \bibinfo {pages} {144101} (\bibinfo {year} {2013})}\BibitemShut {NoStop}%
\bibitem [{\citenamefont {Dong}\ \emph {et~al.}(2023)\citenamefont {Dong}, \citenamefont {Ogunnaike},\ and\ \citenamefont {Levitov}}]{Dong:2023}%
  \BibitemOpen
  \bibfield  {author} {\bibinfo {author} {\bibfnamefont {Z.}~\bibnamefont {Dong}}, \bibinfo {author} {\bibfnamefont {O.}~\bibnamefont {Ogunnaike}},\ and\ \bibinfo {author} {\bibfnamefont {L.}~\bibnamefont {Levitov}},\ }\bibfield  {title} {\bibinfo {title} {Collective excitations in chiral {Stoner} magnets},\ }\href {https://doi.org/10.1103/PhysRevLett.130.206701} {\bibfield  {journal} {\bibinfo  {journal} {Phys. Rev. Lett.}\ }\textbf {\bibinfo {volume} {130}},\ \bibinfo {pages} {206701} (\bibinfo {year} {2023})}\BibitemShut {NoStop}%
\bibitem [{\citenamefont {Pradenas}\ and\ \citenamefont {Tchernyshyov}(2024)}]{Pradenas:2024}%
  \BibitemOpen
  \bibfield  {author} {\bibinfo {author} {\bibfnamefont {B.}~\bibnamefont {Pradenas}}\ and\ \bibinfo {author} {\bibfnamefont {O.}~\bibnamefont {Tchernyshyov}},\ }\bibfield  {title} {\bibinfo {title} {Spin-frame field theory of a three-sublattice antiferromagnet},\ }\href {https://doi.org/10.1103/PhysRevLett.132.096703} {\bibfield  {journal} {\bibinfo  {journal} {Phys. Rev. Lett.}\ }\textbf {\bibinfo {volume} {132}},\ \bibinfo {pages} {096703} (\bibinfo {year} {2024})}\BibitemShut {NoStop}%
\bibitem [{\citenamefont {Dombre}\ and\ \citenamefont {Read}(1989)}]{Dombre:1989}%
  \BibitemOpen
  \bibfield  {author} {\bibinfo {author} {\bibfnamefont {T.}~\bibnamefont {Dombre}}\ and\ \bibinfo {author} {\bibfnamefont {N.}~\bibnamefont {Read}},\ }\bibfield  {title} {\bibinfo {title} {Nonlinear \ensuremath{\sigma} models for triangular quantum antiferromagnets},\ }\href {https://doi.org/10.1103/PhysRevB.39.6797} {\bibfield  {journal} {\bibinfo  {journal} {Phys. Rev. B}\ }\textbf {\bibinfo {volume} {39}},\ \bibinfo {pages} {6797} (\bibinfo {year} {1989})}\BibitemShut {NoStop}%
\bibitem [{\citenamefont {Mermin}\ and\ \citenamefont {Ho}(1976)}]{Mermin:1976}%
  \BibitemOpen
  \bibfield  {author} {\bibinfo {author} {\bibfnamefont {N.~D.}\ \bibnamefont {Mermin}}\ and\ \bibinfo {author} {\bibfnamefont {T.-L.}\ \bibnamefont {Ho}},\ }\bibfield  {title} {\bibinfo {title} {Circulation and angular momentum in the $a$ phase of superfluid helium-3},\ }\href {https://doi.org/10.1103/PhysRevLett.36.594} {\bibfield  {journal} {\bibinfo  {journal} {Phys. Rev. Lett.}\ }\textbf {\bibinfo {volume} {36}},\ \bibinfo {pages} {594} (\bibinfo {year} {1976})}\BibitemShut {NoStop}%
\bibitem [{\citenamefont {Shevelkov}\ \emph {et~al.}(1995)\citenamefont {Shevelkov}, \citenamefont {Dikarev}, \citenamefont {Shpanchenko},\ and\ \citenamefont {Popovkin}}]{Shevelkov:1995}%
  \BibitemOpen
  \bibfield  {author} {\bibinfo {author} {\bibfnamefont {A.~V.}\ \bibnamefont {Shevelkov}}, \bibinfo {author} {\bibfnamefont {E.~V.}\ \bibnamefont {Dikarev}}, \bibinfo {author} {\bibfnamefont {R.~V.}\ \bibnamefont {Shpanchenko}},\ and\ \bibinfo {author} {\bibfnamefont {B.~A.}\ \bibnamefont {Popovkin}},\ }\bibfield  {title} {\bibinfo {title} {Crystal structures of bismuth tellurohalides {BiTe$X$ ($X$ = Cl, Br, I)} from {X-ray} powder diffraction data},\ }\href {https://doi.org/10.1006/jssc.1995.1058} {\bibfield  {journal} {\bibinfo  {journal} {J. Solid State Chem.}\ }\textbf {\bibinfo {volume} {114}},\ \bibinfo {pages} {379} (\bibinfo {year} {1995})}\BibitemShut {NoStop}%
\bibitem [{sup()}]{supmat}%
  \BibitemOpen
  \href@noop {} {}\bibinfo {note} {See Supplementary Material}\BibitemShut {NoStop}%
\end{thebibliography}%

\end{document}